# Modeling the Early Stage Formation Mechanisms of Calcium Silicate Hydrates and Related Gels

Kengran Yang[1,2] and Claire E. White[1,2,*]

[1]Department of Civil and Environmental Engineering, Princeton University, Princeton, NJ, 08544, USA

[2]Andlinger Center for Energy and the Environment, Princeton University, Princeton, NJ, 08544, USA

* Corresponding author: Phone: +1 609 258 6263, Fax: +1 609 258 2799, Email: whitece@princeton.edu

Postal address: Department of Civil and Environmental Engineering, Princeton University, Princeton NJ 08544, USA

**Abstract**

Arguably the most ubiquitous construction material in modern civilization, concrete is enabling the development of megacities around the globe together with increasing living standards in developing nations. However, it can be argued that the concrete industry's somewhat empirical and conservative approach to change is hampering innovation and associated $CO_2$ reductions, yet materials engineering of the essential phase(s) responsible for strength and durability would enable for revolutionary advancements to be achieved. Using a computational materials engineering approach, we simulate the fundamental solution-based building blocks of cement hydrates and their propensity to form larger complexes as assessed from Gibbs free energies of chemical reactions characteristic of the formation of calcium-silicate-hydrate, calcium-alumino-silicate-hydrate and sodium-containing calcium-alumino-silicate-hydrate gels. By accurately simulating the high pH pore solution chemistry in Portland cements and related systems, along with discrete solvation of the species, we discover and discuss the dominant early stage mechanisms controlling gel formation in these systems.

Recent reports from the International Energy Agency (IEA)[1] and the Intergovernmental Panel on Climate Change (IPCC)[2] have highlighted the dire need to substantially curtail greenhouse gas emissions if the world is to limit temperature rise to 2 °C above preindustrial levels. Moreover, as outlined by the IPCC in 2018,[2] if the temperature rise is limited to 1.5 °C this would lead to a more equitable society and increased sustainability yet would require even more drastic changes to greenhouse gas emissions compared with those outlined in previous reports. In the context of construction materials and specifically concrete, ordinary Portland cement (OPC) manufacturing contributes ~5-8% of global $CO_2$ emissions,[3] and various approaches to reduce the net emissions have been outlined by the Cement Sustainability Initiative in conjunction with other organizations.[1] However, for concrete to continue to be used around the globe as developing countries undergo urbanization and other societal changes, drastic improvements in sustainability are required that have yet to be accepted by industry.

The ability to predict a material's performance across length scales would enable for rapid transitions to occur in various industries toward disruptive sustainable technologies. However, within the realm of construction materials, the heterogeneous and multi length scale behavior of the materials[4] has limited our ability so far to use simulations to predict long-term behavior.

Nevertheless, there are key insights that could be gained from accurately simulating the formation of the main phase responsible for strength and durability, namely calcium-silicate-hydrate (C-S-H) gel, where its local bonding environments at the atomic scale can be found in Figure 1, depicted by the crystalline analogue mineral, tobermorite.

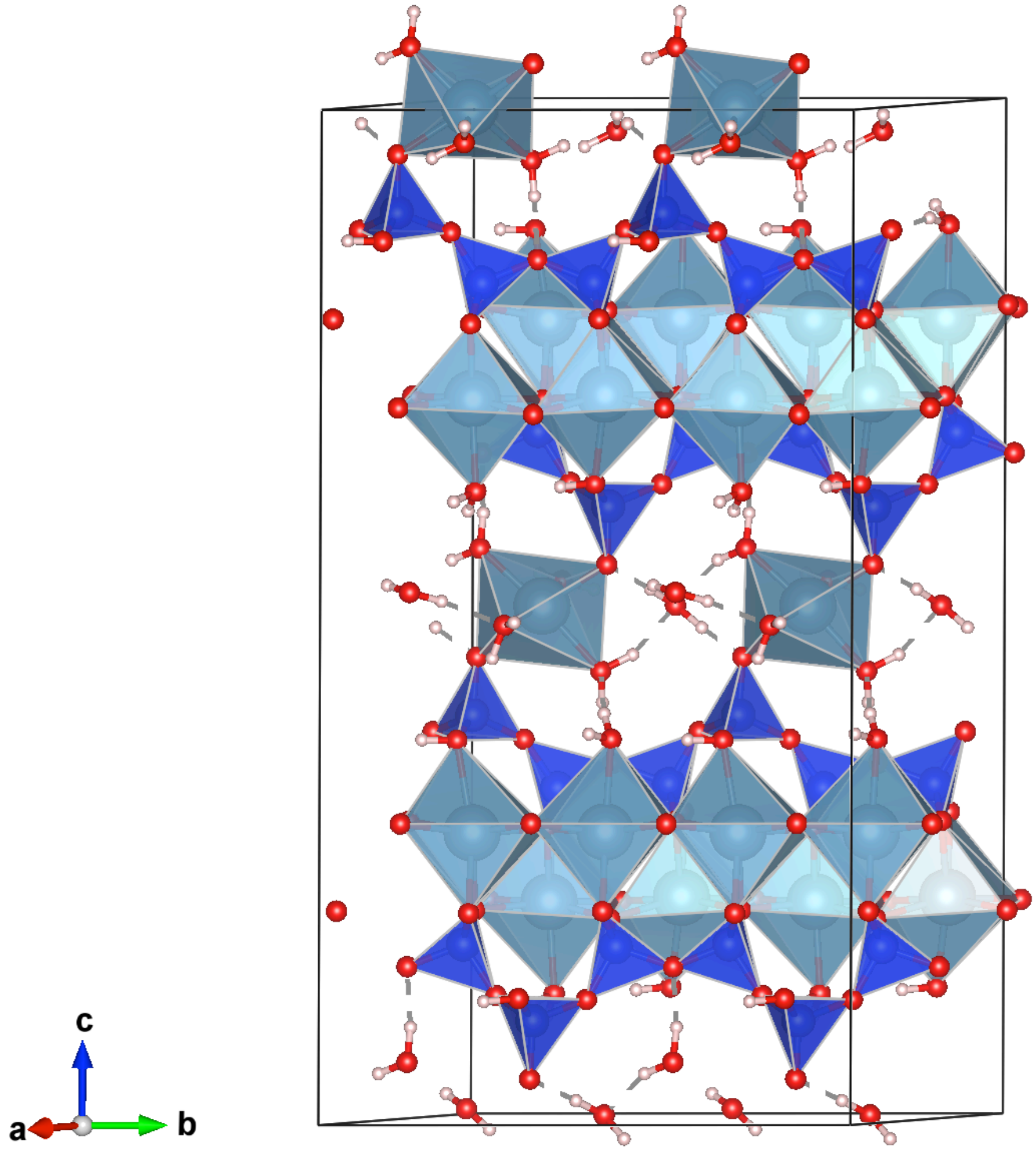

*Figure 1. Structural representation of the local bonding environment in calcium-silicate-hydrate (C-S-H) using the crystalline analogue mineral, tobermorite.[5] Light blue: calcium oxide polyhedra, blue: silicate tetrahedra, red: oxygen atoms, white: hydrogen atoms. Noted that C-S-*

*H has a varied stoichiometry (average of Ca/Si of ~1.7), is amorphous/nanocrystalline, and consists of finite length silicate chains. The tobermorite structure shown here has a Ca/Si ratio of 0.83, is crystalline, and consists of infinite length silicate chains.*

Sustainable disruptive materials that could significantly curtail the greenhouse gas emissions associated with the concrete industry often possess key traits similar to C-S-H gel, but with specific differences in chemistry and/or mesoscale morphology (e.g., pore structure). For example, alkali activation of high-Ca precursors (such as blast furnace slag, class C fly ash, and other Ca-rich materials) leads to the formation of a mechanically-hard binder with mechanical properties equivalent to hydrated OPC. The main binder phase responsible for strength in these alkali-activated materials (AAMs) is sodium-containing calcium-alumino-silicate-hydrate (C-(N)-A-S-H) gel. Hence, by conducting a quantitative assessment of the early stage formation mechanism(s) responsible for the resulting C-S-H and C-(N)-A-S-H gel molecular structures, information on the influence of sodium and aluminum on the resulting structure can be accurately quantified. For example, we recently reported that direct substitution of one calcium atom by sodium and one bridging silicon atom by aluminum in C-S-H (using the model crystalline structure of tobermorite) results in a more stable structure provided the charge deficit is balanced by two hydrogen atoms.[5] Furthermore, given that clinker substitution is being aggressively pursued for OPC, such fundamental studies would provide important insight on how these substitution materials may alter the formation mechanism of C-S-H gel found in OPC systems.

The formation mechanisms of materials are difficult to elucidate using standard experimental characterization techniques, especially for amorphous systems. The manner by which dissolved species in solution rearrange and react largely controls the subsequent growth of specific solid phases. Previous studies on the existence and evolution of such dissolved species include the investigation of calcium carbonate,[6–8] calcium sulfate,[9] aluminosilicates[10,11] and geopolymers.[12,13] Furthermore, beyond the initial reactions between these species, the formation of solid phases and any subsequent phase transformations can have a significant impact on larger length scale properties, such as mechanical and physical attributes. In a recent investigation using transmission electron microscopy (TEM), Schönlein and Plank discovered that amorphous C-S-H globules with sizes of 20 – 60 nm transitioned to larger sized nanocrystalline C-S-H nanofoils during hydration of OPC.[14] Moreover, Krautwurst et al. proposed that formation of C-S-H proceeds via a two-step pathway, where the first step involves formation of amorphous and dispersed spheroids while the second step involves crystallization of tobermorite-type C-S-H from the spheroids, as observed by dynamic light scattering, small-angle X-ray scattering and cryo-TEM.[15] However, the factors controlling the initial formation of the amorphous C-S-H gel have not been studied in the above investigations, specifically which chemical reactions are favored over others, and the influence of sodium and aluminum on these chemical reactions. It is important to note that the growth mechanisms identified in the C-S-H systems studied above were derived from analysis of dilute solutions, and therefore these mechanisms may not be indicative of the complex growth process occurring in hydrated OPC where multiple clinker phases are reacting in a highly concentrated solution.

For decades there have been many investigations studying the nucleation and growth of C-S-H and C-(N)-A-S-H due to their technological importance.[16,17] However, most of these investigations have focused on a much larger length scale than the molecular level,[18] or on doping different cement systems with nucleation agents, such as nanoparticles, which serve as extra nucleation sites for gel growth.[19–22] Moreover, with recent developments in computer simulations, molecular dynamics (MD) has been applied to study the structure of C-S-H and C-(N)-A-S-H and their constituting components (e.g., water, silicate chains, calcium oxide layers) at the nanoscale.[23–32] However, most of these studies[23–28] used a "standard" C-S-H or C-(N)-A-S-H structure, which is based on the structural motifs of tobermorite (crystalline calcium silicate hydrate mineral shown in Figure 1). Therefore, there is an apparent void in the literature on the molecular level interactions between dissolved species in solution that dominate the initial stage of C-S-H and C-(N)-A-S-H formation.[33]

Here, the initial stages of C-S-H, C-A-S-H (calcium-alumino-silicate-hydrate, a variant of C-S-H) and C-(N)-A-S-H formation are investigated using density functional theory (DFT), where all the fundamental ("monomeric") dissolved species (calcium, silicate, aluminate and sodium ions) are explicitly modeled at pH of 12.5 (C-S-H) and 13-14 (C-(N)-A-S-H) using water molecules and a continuum solvation model (Figure 2). The first step of the formation mechanism is explored by calculating the interaction energies (Gibbs free energies) between all possible dissolved species, where the most favorable interactions are identified and correlated with known structural motifs present in the precipitated phases as shown in Figures 3, 4 and 5. Furthermore, by comparing our results with formation energies of C-S-H and C-(N)-A-S-H motifs (e.g., tobermorite-I[34]) used in the literature for thermodynamic modeling, we provide important insight on these literature values that vary up to an order of magnitude.[35–38] Overall, this investigation serves as the very first step to uncovering the formation mechanisms of C-S-H and C-(N)-A-S-H, which are paramount in advancing our understanding of nucleation and growth of cementitious materials and related systems.

## Results

### Solution-based ions and monomeric species

The major clinker phases in OPC powder are tricalcium silicate (or alite, $Ca_3SiO_5$) and dicalcium silicate (or belite, $Ca_2SiO_4$),[39] both of which contain calcium and silicon as major elements. When OPC powder is mixed with water, ions based on silicon and calcium (along with aluminum and iron from additional clinker phases) are released into solution due to dissolution. It is known that at a pH of ~12.5, which is commonly observed in an OPC and water mixture,[40] the major calcium-bearing species are hydrated $Ca^{2+}$ and $CaOH^{+}$,[41] and $Si(OH)_4$, $Si(OH)_3O^{-}$ and $Si(OH)_2O_2^{2-}$ (silicate monomers) for silicon-bearing species.[42] However, it should be noted that the concentration of $Si(OH)_4$ will be minimal at pH > 12 due to the known equilibrium constants of silicate deprotonation.[42] Nevertheless, we include this species here since neutral silicate monomers will likely exist in solution during the very early stages of $Ca_3SiO_5$ and $Ca_2SiO_4$ dissolution (prior to the solution achieving a pH of ~12.5).

OPC can be partially replaced by supplementary cementitious materials (SCMs) such as blast furnace slag (which contains oxides of calcium, magnesium, silicon and aluminum) and class F fly ash (oxides of silicon and aluminum) to improve strength and durability of concrete together with lowering its carbon footprint.[43,44] When these powders dissolve in the high pH environment of hydrating OPC, various ions are released into solution in addition to those based on calcium and silicate. Hence, in this investigation aluminate monomers are also considered, which are in the form of $Al(OH)_4^-$ due to the high pH conditions.[45,46]

For the case of AAMs, ions similar to those in the OPC system are released into solution when a Ca-rich aluminosilicate powder (such as blast furnace slag) is mixed with an alkali source (sodium- and/or potassium-based). Due to the high pH in AAMs (~14),[47] the prevalent calcium species are $CaOH^+$ and $Ca(OH)_2$.[41] Moreover, alkali metal cations from the alkaline solution can charge-balance the electro-negative aluminate/silicate species (i.e., $Si(OH)_3O^-$, $Si(OH)_2O^{2-}$ and $Al(OH)_4^-$) in addition to the charge-balancing behavior of Ca ions. In this study, sodium will be considered as the alkali source (instead of lithium, potassium, rubidium or cesium) since sodium-based alkaline solutions tend to be the most readily available in industry.

Solvation is a key attribute of the dissolved ions and monomers, where each species is solvated by a specific number of water molecules in the first hydration shell. Knowledge of these solvation effects, including the optimal number of water molecules in the first hydration shell, is paramount for accurate determination of reaction energies between the dissolved species.[45] Moreover, it is known that explicit water molecules are needed to improve the accuracy of DFT calculations that use the continuum solvation model (COnductor-like Screening MOdel, COSMO) to replicate the solution environment.[48,49] Due to the difficulty in determining the detailed molecular structure and dynamics of the inner solvation shells of ions/monomers using experimental approaches,[50] a theoretical approach[51] is adopted in this study, as outlined in detail in Supplementary Notes 1-3. With this approach, the optimal number of water molecules for each dissolved species mentioned above has been determined, and their atomic structures are shown in Figure 2.

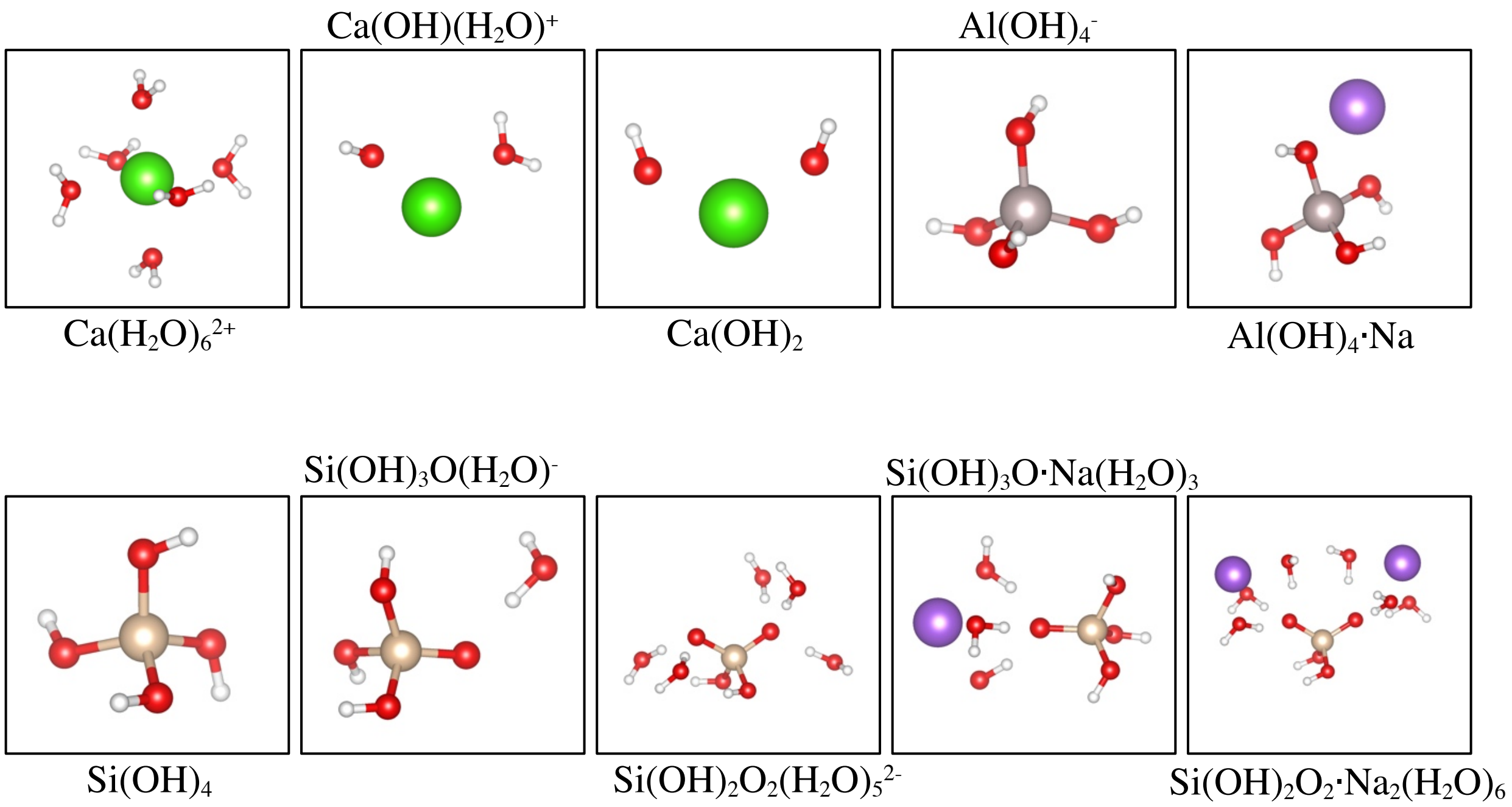

*Figure 2. Atomic structures for the dissolved species involved in OPC and AAM systems together with their optimal number of water molecules. Green denotes calcium, red denotes oxygen, white denotes hydrogen, light grey denotes aluminum, purple denotes sodium, and brown denotes silicon.*

As shown in Figure 2, the optimal number of water molecules for $Ca^{2+}$, $Ca(OH)^+$, $Si(OH)_4$, $Si(OH)_3O^-$, $Si(OH)_2O_2^{2-}$ and $Al(OH)_4^-$ are 6, 1, 0, 1, 5 and 0, respectively. All these species exist in the OPC system ($Al(OH)_4^-$ from supplementary cementitious materials), while $Ca(OH)^+$ and $Si(OH)_4$ are also found in the AAM system. For $Ca(OH)_2$ in AAMs the optimal number of water molecules is 0. As for the silicate species, their optimal number of water molecules have been determined in our previous study, where it was found that 3 water molecules per sodium cation is ideal (i.e., $Si(OH)_4$, $Si(OH)_3O{\cdot}Na(H_2O)_3$, $Si(OH)_2O_2{\cdot}Na_2(H_2O)_6$). Moreover, the same study revealed that the optimal number of water for the aluminate species ($Al(OH)_4{\cdot}Na$) is 0.[10] A detail survey of the literature shows that the optimal number of water molecules calculated here agrees reasonably well with existing results (see Supplementary Note 1 for details). Hence, these results are the first step toward accurate determination of the early-stage formation mechanisms of C-S-H, C-A-S-H and C-(N)-A-S-H.

## Early stage formation mechanisms

Prior to precipitation of the C-S-H phase during OPC hydration, the collective behavior and associated interactions of the dissolved species in solution (outlined in Figure 2) will strongly influence subsequent nucleation and growth of C-S-H. However, at present this behavior remains elusive, preventing researchers from being able to manipulate the early stage formation mechanisms and subsequent C-S-H development. Hence, by uncovering this behavior for OPC and AAM systems, it will then be possible to optimize the chemistry of various cementitious systems to enhance favorable properties (e.g., thermodynamic stability) of the main binder gels. In the following sections the interaction energies of dissolves species for the different gels are reported, specifically for C-S-H, C-A-S-H and C-(N)-A-S-H, together with an in-depth discussion of most probable formation mechanisms (most thermodynamically favorable) for each system based on the calculated thermodynamic data.

### *Calcium-silicate-hydrate (C-S-H)*

Figure 3 shows a schematic representation of the bonding environments found in C-S-H, where key structural components have been identified. C-S-H has an inherent layered structure, consisting of a calcium oxide layer (intralayer calcium sites) situated between silicate chains of finite length.[52] The interlayer space contains variable amounts of calcium ions (referred to as interlayer calcium sites) and water molecules. Our calculations probe the strength of interactions of all dissolved species that ultimately lead to nucleation and growth of C-S-H. We have grouped these interactions by different chemical elements, as outlined in Table 1. This table summarizes the Gibbs free energy of interaction across all relevant "monomeric" species for the OPC system, calculated by the formula $\Delta G_{reaction} = G_{product} - G_{reactant}$. Specifically, the interactions among calcium-bearing species are associative, those between calcium-bearing and aluminate/silicate species are ion-pairing in nature, and reactions among the aluminate and silicate species are condensation reactions, where water molecules are released due to the formation of T-O-T linkages (T denotes tetrahedral Si or Al).

*Table 1. Interaction energies (Gibbs free energies) of the dissolved species in C-S-H system. Values in kJ/mol. Negative values indicate favorable (i.e., spontaneous) reactions.*

| | $Ca(H_2O)_6^{2+}$ | $Ca(OH)(H_2O)^+$ | $Si(OH)_4$ | $Si(OH)_3O(H_2O)^-$ | $Si(OH)_2O_2(H_2O)_5^{2-}$ | $Al(OH)_4^-$ |
|---|---|---|---|---|---|---|
| $Ca(H_2O)_6^{2+}$ | 132.9 | 45.8 | 52.4 | -19.2 | -41.9 | -27.3 |
| $Ca(OH)(H_2O)^+$ | / | -68.9 | 23.8 | -48.6 | -54.0 | -35.1 |
| $Si(OH)_4$ | / | / | -2.3 | -20.2 | -26.3 | -34.9 |
| $Si(OH)_3O(H_2O)^-$ | / | / | / | -10.3 | 30.9 | 3.8 |
| $Si(OH)_2O_2(H_2O)_5^{2-}$ | / | / | / | / | 41.1 | 36.8 |
| $Al(OH)_4^-$ | / | / | / | / | / | 29.6 |

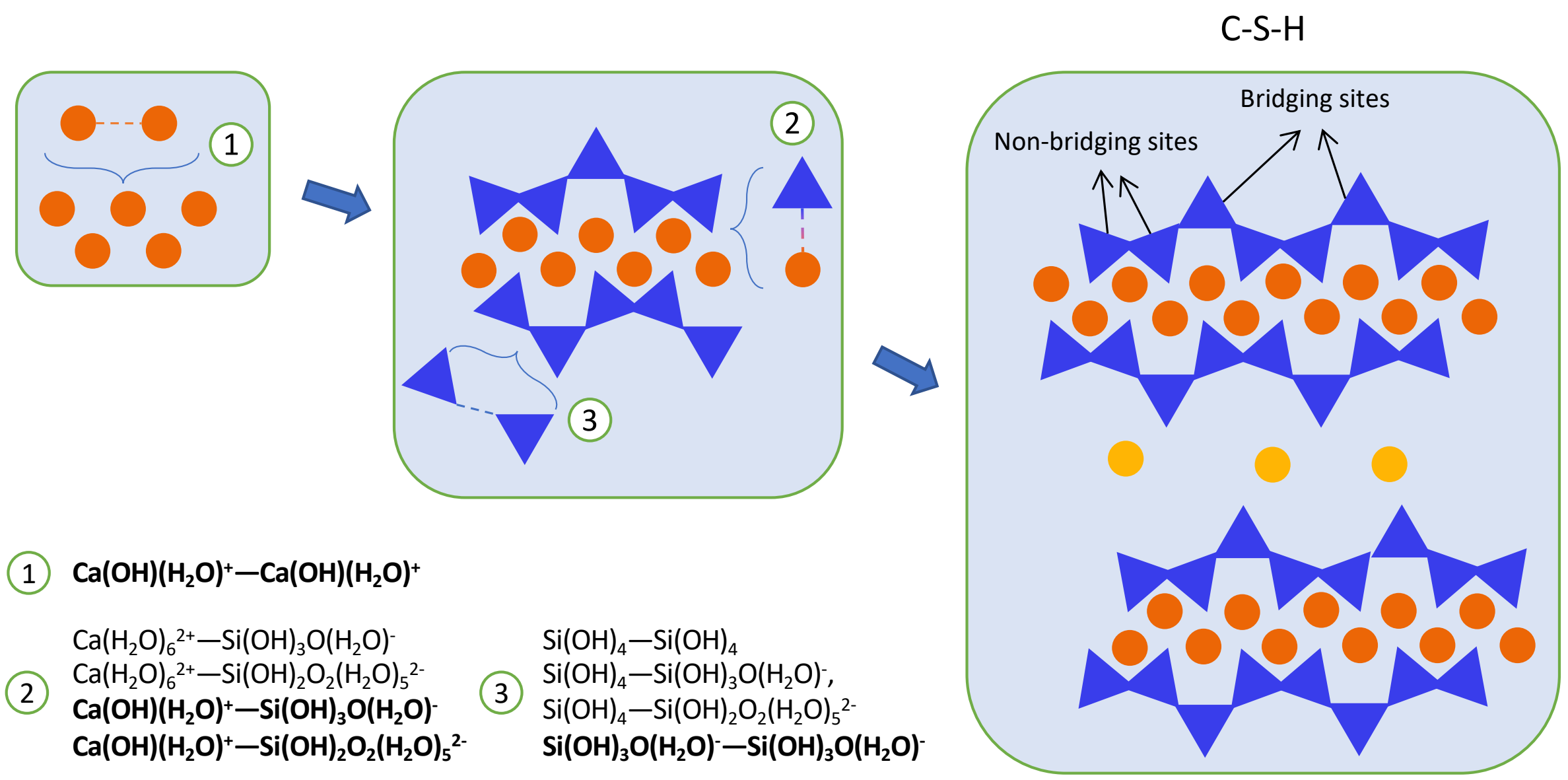

*Figure 3. Schematic representation of the bonding environments in C-S-H and the early stage formation mechanisms involving select interactions between calcium- and silicon-bearing species identified by DFT calculations. The resulting structural motifs found in the C-S-H structure are also highlighted. Bolded reactions are considered more likely to happen for the given reaction type (see text for details). The numbers indicate specific reaction steps of the formation mechanism. Blue triangles, orange dots and yellow dots represent tetrahedral silicate, intralayer calcium and interlayer calcium, respectively. Note that the oxygen atoms in the calcium oxide layer are not shown for clarity.*

The most favorable interaction in Table 1 is that between two calcium monohydroxide species, $Ca(OH)(H_2O)^+$-$Ca(OH)(H_2O)^+$, with a ΔG value of -68.9 kJ/mol. Hence, as OPC powder dissolves and releases calcium and silicate (and other) ions into solution, the calcium monohydroxide ions will quickly associate, with this early stage formation reaction likely leading to the formation of the calcium oxide layer motif in C-S-H (Step 1 in Figure 3). Furthermore, Table 1 shows that only favorable reaction involving calcium is the interaction between two $Ca(OH)(H_2O)^+$ complexes. The other reactions are unfavorable due to the large electrostatic repulsion when a $Ca(H_2O)_6^{2+}$ cluster is present (132.9 and 45.8 kJ/mol for $Ca(H_2O)_6^{2+}$ interacting with $Ca(H_2O)_6^{2+}$ and $Ca(OH)(H_2O)^+$, respectively).

The second most favorable interaction in Table 1 is between $Ca(OH)(H_2O)^+$ and $Si(OH)_2O_2(H_2O)_5^{2-}$ with a $\Delta G$ value of -54.0 kJ/mol. In fact, it can be seen in Table 1 that all interactions between calcium-bearing species ($Ca(H_2O)_6^{2+}$ and $Ca(OH)(H_2O)^+$) and negatively charged silicate species ($Si(OH)_3O(H_2O)^-$ and $Si(OH)_2O_2(H_2O)_5^{2-}$) are thermodynamically favorable. On the other hand, the dimerization (i.e., condensation) reactions involving silicate species are less favorable (largest magnitude $\Delta G$ of -26.3 kJ/mol for $Si(OH)_4$-$Si(OH)_2O_2(H_2O)_5^{2-}$). Hence, these results provide important mechanistic insight on how the silicate monomers behave in the OPC alkaline solution (high pH). Instead of the silicate monomers undergoing oligomerization in solution (as is the case for aluminosilicate solutions),[53] they will initially be subjected to strong ion-pairing with calcium-bearing species (Step 2 in Figure 3). Such calcium-silicate structural motifs are found in C-S-H, specifically between the silicate chains and intralayer calcium together with bridging silicates and interlayer calcium. The unfavorable interactions of $Ca(H_2O)_6^{2+}$-$Si(OH)_4$ and $Ca(OH)(H_2O)^+$-$Si(OH)_4$ (Table 1, 52.4 and 23.8 kJ/mol, respectively) are due to the repulsive electrostatic forces between the species. However, since the presence of neutral silicate will be minimal in such high pH (> 12) environments (due to the rapid congruent dissolution of tricalcium silicate in water and subsequent rise in pH[54]),[42] these repulsive interactions will not adversely impact the formation mechanisms during C-S-H nucleation and growth.

The existence of $Ca(OH)^+$ in the C-S-H structure is supported by the experimental finding from inelastic neutron scattering, where ~23% of calcium in C-S-H gel (with Ca/Si = 1.7) is charge-balanced by -OH groups.[55] However, it should be noted that the fraction of $Ca(OH)^+$ in the mixture of OPC and water is relatively small (~7%) compared to the $Ca^{2+}$ species (~93%) at pH of ~12.5,[41] implying that the growth rate of the calcium oxide layer is likely controlled by the conversion rate of $Ca(OH)^+$ from $Ca^{2+}$ once the small amount of $Ca(OH)^+$ has been taken out of solution. Although $Ca^{2+}$ species may not directly participate in the formation of the calcium oxide layer, these species may reside in the interlayer spacing of C-S-H (yellow dots in Figure 3), or interact with $Si(OH)_3O(H_2O)^-$ and $Si(OH)_2O_2(H_2O)_5^{2-}$ (as indicated by Table 1) to form intermediate complexes.

The formation energies of key reactions between calcium- and silicon-bearing species at standard state (25°C and 1 atm) can be found in thermodynamic databases. For OPC hydration, Lothenbach and Winnefeld[34] reported that the formation energies of $CaSi(OH)_2O_2$ and $CaSi(OH)_3O^+$ clusters are -25 and -7 kJ/mol, respectively, as determined by potentiometric titrations in a mixture of $Ca(ClO)_2$, $SiO_2$, NaOH and $NaClO_4$.[56,57] In contrast, the formation energies of these two clusters calculated here are -42 and -19 kJ/mol, respectively. Hence, although there are differences in magnitude, the overall trend is present, specifically that $CaSi(OH)_2O_2$ formation is more favorable than $CaSi(OH)_3O^+$. Moreover, the formation energy of $CaSi(OH)_2O_2$ (without any solvating water molecules) has also been calculated by Galmarini et al. using metadynamics in force field MD, with a result of -57 kJ/mol.[58] Since our result is closer to the experimental values reported by Lothenbach and Winnefeld compared with those of Galmarini et al., this further confirms the suitability of the BLYP+DNP level of DFT calculation used for these aqueous species.

Key details on the early stage formation mechanism of the silicate chains in C-S-H can be drawn from the interaction energetics between silicate species in Table 1, which is also summarized in Step 3 of Figure 3. It is clear that the condensation reactions involving neutral silicate ($Si(OH)_4$-$Si(OH)_4$, $Si(OH)_4$-$Si(OH)_3O(H_2O)^-$, $Si(OH)_4$-$Si(OH)_2O_2(H_2O)_5^{2-}$) are all favorable (-2.3, -20.2 and -26.3 kJ/mol, respectively), together with the interaction between two singly deprotonated silicates ($Si(OH)_3O(H_2O)^-$-$Si(OH)_3O(H_2O)^-$, -10.3 kJ/mol). On the other hand, the reactions involving doubly deprotonated silicate ($Si(OH)_3O(H_2O)^-$-$Si(OH)_2O_2(H_2O)_5^{2-}$ and $Si(OH)_2O_2(H_2O)_5^{2-}$-$Si(OH)_2O_2(H_2O)_5^{2-}$) are thermodynamically unfavorable (30.9 and 41.1 kJ/mol, respectively) due to the strong electrostatic repulsion between two negatively charged species.[59] Hence, these quantitative results on silicate dimerization (in comparison with ion-pairing between calcium and silicates) indicate that the formation of the silicate chains is instigated by the dominant behavior of the calcium-calcium association interactions together with calcium-silicate ion pairing reactions due to their higher thermodynamic favorability, and therefore the calcium species are largely driving the early stage formation mechanism of C-S-H in OPC systems. Discussion of silicate dimerization in the context of existing literature can be found in Supplementary Note 4.

*Calcium-alumino-silicate-hydrate (C-A-S-H)*

When blast furnace slag (~15 wt. % $Al_2O_3$[44]) or coal-derived fly ash (~20 – 40 wt. % $Al_2O_3$[60]) are used as SCMs in OPC concrete, aluminate species are present in the system and are known to partially replace silicate species in C-S-H (Figure 4), leading to the formation of calcium-alumino-silicate-hydrate (C-A-S-H). Hence, it is important to study the influence of alumina on the dimerization and ion-pairing reactions, including whether alumina augments the early age formation mechanisms discussed above for C-S-H. Table 1 shows that the $Al(OH)_4^-$ monomer prefers to bond with more electronically positive species, with favorable reactions for $Ca(H_2O)_6^{2+}$-$Al(OH)_4^-$, $Ca(OH)(H_2O)^+$-$Al(OH)_4^-$ and $Si(OH)_4$-$Al(OH)_4^-$ species (-27.3, -35.1 and -34.9 kJ/mol, respectively), and unfavorable reactions for $Si(OH)_3O(H_2O)^-$-$Al(OH)_4^-$, $Si(OH)_2O_2(H_2O)_5^{2-}$-$Al(OH)_4^-$ and $Al(OH)_4^-$-$Al(OH)_4^-$ species (3.8, 36.8 and 29.6 kJ/mol, respectively). Hence, as was the case for ion-pairing between calcium and silicate species, there is strong ion-pairing between calcium and aluminate. However, in contrast to C-S-H, where relatively weak dimerization reactions were calculated (maximum value of -26.3 kJ/mol), the dimerization reaction in C-A-S-H involving an aluminate monomer and neutral silicate monomer is slightly more favorable (-34.9 kJ/mol). Nevertheless, due to the absence of these silicate monomers at high pH, the early stage formation mechanism involving aluminate will be dominated by calcium-aluminate of ion-pairing. There is a complete deficit of information in the literature regarding the interaction energies between calcium and aluminate species, and only a limited amount on the interaction between silicate and aluminate monomers, as discussed in Supplementary Note 5.

It is known from NMR experiments that aluminate is found in C-A-S-H on the bridging sites in the silicate chains[61], yet from our calculations it is unclear how the early stage formation mechanisms lead to the aluminate species residing on these sites. It is possible that the aluminate species initially resides on a non-bridging site (see Figure 3 for location of non-bridging site) due to the strong interactions between calcium species forming the calcium oxide layer and the strong ion-pairing interactions between calcium and aluminate. However, due to thermodynamic interactions (such as multi-body and long-range interactions) not taken into account in this study, the partially formed silicate chains react readily with the aluminate species, leading to the aluminate moving to a bridging site. On the other hand, another possible explanation is that the aluminate species in an ion-pair ($Ca(OH)(H_2O)^+$-$Al(OH)_4^-$) is initially associated with the calcium in the interlayer (yellow dots in Figure 4), which would directly lead to the aluminate existing on a bridging site (shown in Figure 4).

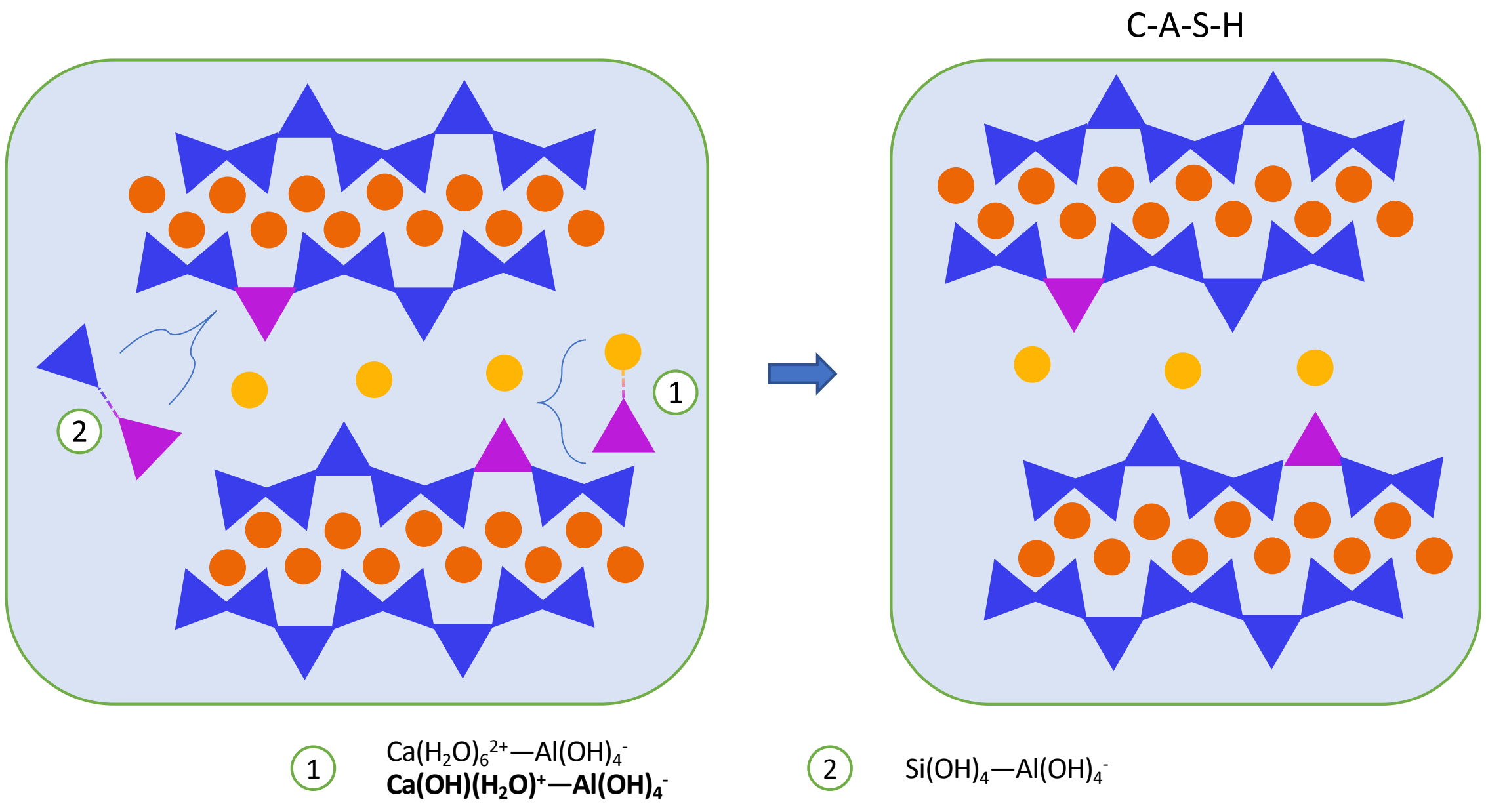

*Figure 4. Schematic representation of the bonding environments in C-A-S-H and the early stage*

*formation mechanisms involving select interactions between calcium-, silicon- and aluminum-bearing species according to our calculations. Bolded reactions are considered more likely to happen for the given reaction type (see text for details). The numbers indicate specific reaction steps of the formation mechanism. Blue triangles, orange dots and yellow dots represent tetrahedral silicate, intralayer calcium and interlayer calcium, respectively. Purple triangles represent tetrahedral aluminate. Note that the oxygen atoms in the calcium oxide layer are not shown for clarity.*

### *Sodium-based calcium aluminosilicate hydrate (C-(N)-A-S-H)*

In AAMs, the bonding environments of the major binding phase (C-(N)-A-S-H, Figure 5) are different from those of C-A-S-H (Figure 4) due to the higher pH environment and existence of sodium cations. The influence of sodium on the interactions between dissolved species is given in Table 2. Due to the high pH (~14) environment common in systems where C-(N)-A-S-H gel forms, the most prevalent calcium species are $Ca(OH)(H_2O)^+$ (singly charged) and $Ca(OH)_2$ (neutral), making up ~20% and ~70% of the total calcium-bearing species, respectively.[41] As evident in Table 2, the strongest interactions are amongst calcium-bearing species (most favorable reaction of -83.2 kJ/mol for $Ca(OH)(H_2O)^+$-$Ca(OH)_2$), with all Ca-Ca association interactions being thermodynamically favorable. Therefore, there are more pathways to form the calcium oxide layer in the C-(N)-A-S-H system compared to C-S-H system (where only the $Ca(OH)(H_2O)^+$-$Ca(OH)(H_2O)^+$ is favorable at -68.9 kJ/mol), and a stronger driving force exists for this formation to occur due to the magnitude of the favorable reactions. Furthermore, the stronger interactions for Ca-Ca species in Table 2 compared with Table 1 is in agreement with the experimental observation that the solubility of $Ca(OH)_2$ (i.e., portlandite) in water markedly decreases with addition of NaOH.[62]

*Table 2. Interaction energies of the dissolved species in C-(N)-A-S-H system. Values in kJ/mol. Negative values indicate favorable (i.e., spontaneous) reactions.*

| | $Ca(OH)(H_2O)^+$ | $Ca(OH)_2$ | $Si(OH)_4$ | $Si(OH)_3ONa(H_2O)_3$ | $Si(OH)_2O_2Na_2(H_2O)_6$ | $Al(OH)_4Na$ |
|---|---|---|---|---|---|---|
| $Ca(OH)(H_2O)^+$ | -68.9 | -83.2 | 23.8 | -17.6 | -38.3 | -44.5 |
| $Ca(OH)_2$ | / | -74.3 | -31.9 | -54.1 | -54.2 | -33.3 |
| $Si(OH)_4$ | / | / | -2.3 | -3.4 | -17.1 | -19.9 |
| $Si(OH)_3ONa(H_2O)_3$ | / | / | / | 5.6 | 8.5 | -9.9 |
| $Si(OH)_2O_2Na_2(H_2O)_6$ | / | / | / | / | 27.6 | 4.9 |
| $Al(OH)_4Na$ | / | / | / | / | / | 8.9 |

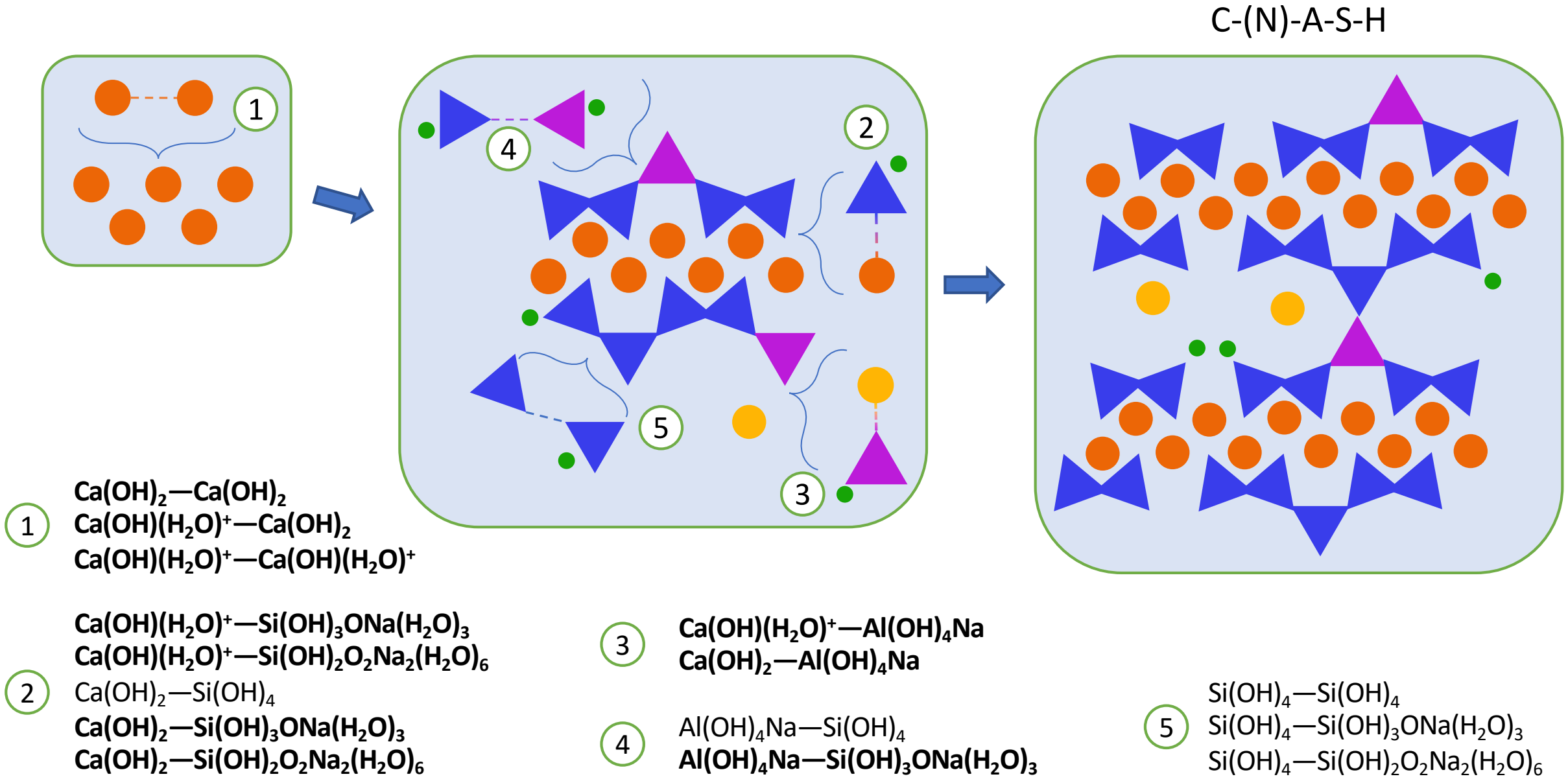

*Figure 5. Schematic representation of the bonding environments in C-(N)-A-S-H and the early stage formation mechanisms involving select between calcium-, silicon-, aluminum-bearing species (including charge-balancing sodium ions) according to our calculations. Bolded reactions are considered more likely to happen for the given reaction (see text for details). The numbers indicate specific reaction steps of the formation mechanism. Blue triangles, orange dots and yellow dots represent tetrahedral silicate, intralayer calcium and interlayer calcium, respectively. Purple triangles represent tetrahedral aluminate and green dots are the sodium ions. Note that the oxygen atoms in the calcium oxide layer are not shown for clarity.*

Table 2 shows that the next type of interactions that are most favorable are the ion-pairing reactions between calcium and aluminate/silicate species, especially those between $Ca(OH)_2$ and singly- or doubly-deprotonated silicate (-54.1 and -54.2 kJ/mol, respectively). Hence, as was the case for C-S-H and C-A-S-H, these early stage formation mechanisms in C-(N)-A-S-H likely result in the calcium oxide layer motifs with aluminate/silicate species being actively incorporated into the layered structure via strong ion-pairing interactions, subsequently forming the aluminosilicate chains.

Since $Ca(OH)(H_2O)^+$ exists in C-(N)-A-S-H, C-A-S-H and C-S-H systems, its interaction with the aluminate/silicate species can be used to examine the impact of sodium on the early stage formation mechanisms. It can be seen that the formation of $Ca(OH)(H_2O)^+$-$Si(OH)_3ONa(H_2O)_3$ and $Ca(OH)(H_2O)^+$-$Si(OH)_2O_2Na_2(H_2O)_6$ species becomes less favorable due to the presence of sodium (initially -48.6 and -54.0 kJ/mol, respectively, to -17.6 and -38.3 kJ/mol when sodium is present), while the interaction between $Ca(OH)(H_2O)^+$ and $Al(OH)_4Na$ becomes more favorable (-35.1 kJ/mol without sodium, -44.5 kJ/mol with sodium). Hence, sodium enhances the incorporation of aluminate into the C-A-S-H gel by encouraging ion-pairing between calcium species and the aluminate monomer. This computational finding is supported by existing literature, specifically the study by L'Hopital et al.[63] that investigated the impact of potassium on the resulting Al/Si ratio of C-A-S-H using NMR and other experimental techniques, where a direct correlation between the presence of alkalis and a higher Al/Si for the C-A-S-H was found.

It is clear from comparison of Tables 1 and 2 that the absolute value of the silicate dimerization formation energies decreases with the presence of sodium, indicating that these interactions are weaker due to the reduced charge difference. Kinrade and Pole studied the effects of alkali metal cations on the chemistry of aqueous silicate solutions with $Si^{29}$ NMR, and showed that the alkaline cation promotes the interaction of two negatively charged silicate anions by overcoming their electrostatic repulsion, however, the cation is also seen to hinder the subsequent formation of a siloxane bond.[59] Hence, our DFT observation of weaker interactions between silicate species due to the presence of sodium is in agreement with these experimental findings. It is important to note that the energy values obtained in this study (Table 2) are slightly different from those we obtained in ref. [10] due to a different level of theory being used previously for calculation of the Gibbs free energy of the water molecule associated with the dimerization reaction.

Finally, the introduction of sodium is seen to have a similar effect on the energetics of the dimerization reactions involving aluminate monomers compared with the silicate-silicate dimerization reactions, specifically that the interactions weaken. The only anomaly is the interaction between the $Al(OH)_4Na$ and $Si(OH)_3ONa(H_2O)_3$ monomers, where it becomes favorable after the introduction of sodium to the system. This is supported by the study of Yang et al., where they reported a value of -42 kJ/mol for a slightly different reaction (the silicate monomer is $Si(OH)_3ONa$ instead of $Si(OH)_3ONa(H_2O)_3$) using the same methodology as our study (BLYP+DNP level of theory + COSMO).[11] The significant difference of energy values (-10 kJ/mol here vs. -42 kJ/mol by Yang et al.) may be attributed to the absence of explicit water molecules for the $Si(OH)_3ONa$ cluster in the study by Yang et al., and that their solvent was treated as neutral pH water instead of a high pH solution. Yang et al. also calculated the formation energies of $Al(OH)_4Na$-$Si(OH)_4$ and $Al(OH)_4Na$-$Al(OH)_4Na$ dimers, with values of -21 and -16 kJ/mol, respectively.[64] In contrast, our results for these two clusters are -20 and 9 kJ/mol, respectively. However, given the Lowenstein avoidance rule for charge-balanced Al-O-Al linkages,[65] our unfavorable value of 9 kJ/mol may be more realistic.

## Discussion

Given the limitations of experimental approaches for determining solution speciation and subsequent early stage chemical reactions of cementitious materials at the atomic length scale, there are few relevant studies available in the literature for direct comparison with our proposed formation mechanism. However, Gartner et al. proposed that growth of C-S-H involves adsorption of hydrated calcium ions on the C-S-H basal sheets (i.e., silicate chain – CaO layer – silicate chain structure), where the initial formation of the basal sheets was not discussed.[66] Our study sheds light on the plausible routes of formation of these basal sheets, demonstrating the need and power of atomistic modeling for tackling these fundamental questions at the atomic length scale.

Our results highlight important formation differences between the traditional OPC system and the more sustainable OPC-SCM and AAM systems, which will aid future optimization of these systems and other relevant materials such as zeolites and glasses at the atomic level. However, it is important to note that this study focused on the thermodynamics of the early stage formation mechanism (Tables 1 and 2), and therefore the kinetics of these reactions need to be thoroughly explored as the next stage of research to verify the formation mechanisms of calcium silicate hydrates reported here.

In conclusion, this investigation has elucidated the early stage formation mechanisms of C-S-H, C-A-S-H and C-(N)-A-S-H gel using density functional theory calculations. By modeling the solution speciation found in OPC (specifically tricalcium silicate), OPC containing Al-rich SCMs, and high-Ca AAMs, the Gibbs free energies of all possible interactions between the solution species were determined. Irrespective of cement system, the calcium-calcium interactions were seen to be the most favorable reactions, indicating that formation of these gels is initiated by formation of the calcium oxide layer motif. Inclusion of silicates (and aluminates) in the gel will be dominated by the strong ion-pairing reactions between calcium and the silicate-bearing species, since these ion-pairing reactions were found to be the second most dominant reactions (second largest in magnitude) found in the cement systems. Interestingly, the inclusion of sodium in the ion-pairing reactions was found to strengthen the calcium-aluminate interaction, and therefore incorporation of aluminates into the C-S-H gel is enhanced by the presence of sodium (as is the case for high-Ca AAMs). The weakest interaction type was found to be those between silicate-silicate, and silicate-aluminate species (i.e., condensation reactions), irrespective of the pH (12.5 of 14) and presence of sodium. Hence, the early formation mechanism of C-S-H-type gels appears to be dominated by the calcium-containing interactions, with the evolution of the silicate chains being a direct consequence of the strong calcium-calcium and calcium-silicate interactions.

## Methods

### DFT calculation

Density functional modeling was carried out with the exchange-correlation potential approximated by the BLYP functional. The BLYP functional has been widely used for aluminosilicate systems[10,67] and aqueous ionic solutions.[68] Recently, it was used for the C-S-H system and proved to be effective.[24] The basis set adopted was the double numerical plus a polarization p-function on all hydrogen atoms (DNP) to account for hydrogen bonding, which has a better performance per computational cost compared to the Gaussian-type basis sets.[69] No pseudopotentials or effective core potentials were used. A self-consistent field (SCF) convergence of $10^{-7}$ hartrees was used with no smearing, along with an orbital cutoff of 8.0 Å for all atom types. The solution environment was simulated using the continuum solvation model (COnductor-like Screening MOdel, COSMO) together with the inclusion of explicit water molecules.[51] A dielectric constant of 56 was used in the COSMO model to mimic a high pH environment, as the dielectric constant is measured to be 56 for a NaOH solution with a concentration of 2 mol/L, which has a pH of about 14.[70]

Three major tasks were performed in this study: geometry optimization and free energy calculations which were run on a single core using DMol$^3$ v4.4 package, and ab initio molecular dynamics (MD) which was carried out using 8 – 16 cores and the DMol$^3$ v7.0 package. To obtain a proper starting structure as an input for the geometry optimization, simulated annealing was performed with ab initio MD at different temperatures. The highest annealing temperature was determined such that the cluster remained intact (i.e., atoms do not dissociate from each other) during simulation at that given temperature. Simulations were run for 2 – 5 ps at the highest annealing temperature to allow for sufficient exploration of the geometrical configuration of the species, with a time step of 1 fs. A number of starting structures were obtained by finding the minima in the potential energy profile of the species during the MD run. To save computational cost, the PWC functional was used, together with the DNP basis set. Simulations were conducted using the NVT ensemble, with the temperature being controlled by a Nose-Hoover thermostat.

Geometry optimization convergence thresholds were set at $1 \times 10^{-6}$ hartrees for energy, $2 \times 10^{-4}$ hartrees/Å for maximum force, and $5 \times 10^{-4}$Å for maximum displacement for most calculations, though sometimes more stringent thresholds were used due to the flat energy landscape of clusters with a sizable amount of hydrogen bonding.[68] Vibrational frequency analysis was performed on each geometry-optimized structure to ensure that it was located at a local minimum on the energy landscape, as well as to obtain its Gibbs free energy at 298.15K. The process has been outlined in our previous study.[10] All analysis was carried out using the Accelrys Materials Studio software.

It is important to note that the values in Table 1 and 2 may not represent the true global minimums on the energy landscape, the search of which has been a long-standing issue in quantum chemical calculations.[71] Energy values from different local minima for each reaction are reported in Supplementary Note 6, where non-negligible energy variations are observed. Due to these variations, an approximate standard deviation of $\pm 10.3\ kJ/mol$ is suggested when assessing the energy values in Tables 1 and 2 without supporting theoretical or experimental data.

## Acknowledgments

This work was supported by NSF through the MRSEC Center (Grant No. DMR-1420541). The calculations presented in this article were performed on computational resources supported by the Princeton Institute for Computational Science and Engineering (PICSciE) and the Office of Information Technology's High Performance Computing Center and Visualization Laboratory at Princeton University.